# THE "SUN - CLIMATE" RELATIONSHIP : III. THE SOLAR ERRUPTIONS, NORTH-SOUTH SUNSPOT ARREA ASSYMETRY AND CLIMATE

Boris Komitov

Bulgarian Academy of Sciences- Institute of Astronomy, 6003 Stara Zagora-3, POBox 39,
b_komitov@sz.inetg.beg

In this last Paper III additional evidences that the solar high energetic particles radiation with energies higher as 100 MeV (the solar cosmic rays SCR) is an very important component for the "Sun- climate" relationship are given (see also Paper I and II). The total solar irradiance (TSI) and the galactic cosmic rays (GCR) variations given an integral climate effect of cooling in sunspot minima and warming in the sunspot maxima. Unlike the both ones the powerful solar corpuscular events plays a cooling climate role during the epochs of their heigh levels. By this one subcenturial global and regional temperature quasi- cyclic changes by duration of approximately 60 years could be track during the last 150 years of instrumental climate observations . It has been also evided in the paper that this subcenturial oscilation is very important in the Group sunspot number (GSN) data series since the Maunder minimum up to the end of 20$^{th}$ century. Thus the solar erruptive activity effect make the total "Sun –climate" relationship essentially more complicated as it could be follow when only the TSI and GCR variations are taken into account. In this light the climate warming tendency after AD 1975 is rather by a natural as by an antropogenic origin. Most probably the last one is very close related to the general downward tendency of erruptive solar events which is superimposed over the high long term TSI levels during the last three decades (AD 1975-2007).

It is evided, that the efficiency of the solar corpuscular activvuty over the climate is strongly depended by the "north-south" assymetry of the solar activity centers (as a proxy the sunspots area north-south assymetry index $A$ is used there). The climate cooling effect in the Northern hemisphere is most powerful during the epochs of positive values of $A$. This effect is very significant in combination with high level of the GSN-index . A strong quasi 120 year "hypercycle" has been detected in the $A$ index during the period of AD 1821-1994. Most probably the observed 120-130 yr cyclity in climate and $^{10}$Be continental ice core data (both "Greenland" and "Antarctic" series) is related to the last one.

The expected climate changes during the next decades and especially during the new solar sunspot cycle No 24 are discussed on the base of the "multiple" nature of the "Sun-climate" relationship.

*1.Introduction*

According the most perceived point of view the "Sun-climate" relationship during the present postglacial era (Holocene, the historical time scale) is realized predominantly by the total solar irradiance (TSI) variations (Solanki , 2002; de Jaeger and Usoskin, 2006 ). The TSI index is well known since AD 1978 on the base of satellite observations (Frolich et al, 1997; Pap et al. , 2003 ) The last one during the last ~ 400 years corresponds very well to the overall sunspot activity (the International Wolf's number $Ri$ and the Group sunspot number ($GSN$ or $Rh$)( Lean et al., 1995; Lean, 2000, 2004). There are also a significant number of theoretical (numerical) , mixed type (statistical + theoretical) or "poor" statistical studies in which the relationship "sunspot activity - solar magnetic flux -> TSI" is investigated (Lean et al,2000; Solanki et al, 2002; Krivova et al., 2007 etc.).

On other hand there are evidences that an additional mechanism of indirect Sun' s forcing over the climate due to the modulation of galactic cosmic rays (GCR) by solar wind exist. The first works in this course are still from the middle of 1970$^{st}$ (Dickinson, 1975). The aerosols and clouds production rates forcing in the lower atmosphere under the GCR-flux increasing during the sunspots minimuma epochs is discussed by Svenmark and Friiz-Christiensen (1997) and Yu (2002) . There are also some interesting results of Tinsley (2000), concerning the GCR-flux influence over the atmospheric electricity and circulation.

It has been marked by many authors that the "overall sunspot activity - TSI -> climate" relationship is far not enough to explain the real climate dynamics during the last 400 years since AD 1610 . As it is pointed out by Thompson (1997) only 25% of the global warming effect after AD 1850 could be explained due to the TSI increasing during the this time. For the other 75% should be search for additional factors. Especially after AD 1975/80 there is a total divergence between the TSI and global temperature changes (Solanki, 2002; Usoskin et al.,2005; Lockwood and Frolich, 2007). The phenomena couldn't to explain satisfactory even if in addition the GCR-flux is taken into account. This is why for the last 30-35 years by the opinion of many researchers the human activity is the factor , which play the dominant role for the climate changes.

It has been shown in the first paper of this series (see Paper I), that the residual variations to the regressional models "sunspot activity – temperature data" both for the Northern hemisphere (AD 1610 -1979) (Moberg et al., 2005) and for the World Ocean (1856-1995) (Parker et al., 1995) are far not occasional. There are well expressed cyclic oscilations in the quasi-centurial and subcenturial range. The spectra of the last ones is more complicated in the Northern hemisphere "residual" data series ( powerful cycles by duration of 54-67 (doublet) and 120-130 years), while in the World Ocean one there is only a strong cyclic 58-63 year oscilation (doublet) as well as essentially weaker trace of 88 year one. It has been summarized finally in Paper I that there is powerful quasi- 60 year climatic cycle in the modern epoch . The last one plays a very important role in the climate , causing few waves of cooling and warming since the end of Maunder minimum , which are superimposed over the general regressional relationship "sunspot activity –temperature data" during this time. It has been also shown in Paper I that the climate warming epoch after AD 1975 up to 2005-2006 well correspond to the serial upward phase of this 60 year cycle.

It has been found in the next Paper II that a very powerful quasi –60 year cycle exist both in the middle latitude aurora (MLA) (Krivsky and Pejml, 1988) as well as in the "Greenland" $^{10}$Be data series (Beer et al., 1990,1998). It has been shown that there is a very good coincidience by time between the corresponding 60-year cycle extremums in the both series. The local 60 yr maximums in MLA and $^{10}$Be series during the last 300 years since AD 1700 correspond to subcenturial temperature minimums . They are well expressed in the both studied temperature series, but essentially better in the World Ocean ones.

The MLA events occurs in the upper Earth atmosphere and by this one they are strongly independent from the troposphere processes and the climate. Their primary sources are active events such as the coronal mass ejections (CME). Consequently this is related to the 60 year cycle in this series too. Due to this fact it has been concluded in Paper II that : 1 The quasi –60 year $^{10}$Be cycle is most probably by solar origin and it is caused by an yield of solar high energetic protons E> 100 MeV in the total production rate of this "cosmogenic" isotope in the stratosphere; 2. The quasi 60 year climate cycle is caused by the same one high energetic solar corpuscular events; 3. The increasing of the solar high energetic particles ( probably mainly protons) fluxes lead to the same effects in Earth atmosphere such as the galactic cosmic rays with the same energies, i.e. an increasing of the aerosols production rate and cloudness and as a final effct – to a climate cooling.

The aim of this last paper (Paper III) is to make a more detailed analysis and to give additional arguments for the important role of erruptive solar processes as a climate forcing factor. A strong evidence that a very important role there the north-south assymetry of the erruptive events is also played is given.

Evidences in this course that the modern climate changes are not by antropogenic origin could also given by comparuson of Mars climate conditions during the last three decades. It has been found that the total albedo of this planet has been falled down from AD 1977 up to 1999 , which correspond to a warming in order of 0.6 K only for about of two decades ().

It is demonstrated that the total "Sun-climate" relationship is much more complicated as it is follows if only the TSI and GCR flux changes are taken into account. However, in the same time it is much better fit to the real observed climate variations both in the presence and in the past.

*2. Data and Methods*

The following data sets are used in this study:
- The Northern hemishpere temperature data series from AD 1610 to 1979 – the last 370 years from the data set of Moberg et al. (2005). As a "zero level" the mean temperature between AD 1961 and 1990) is used there
- The World Ocean temperature data series from AD 1856 to 1995 (Parker et al.,1995). The "zero level" there is the mean temperature in AD 1940.
- The middle latitude aurora (MLA) annual number data series from AD 1700 to 1900 , i.e. the last two centuries from the catalogue of Krivsky and Pejml (1988) with most certain data. The catalogue data for the period AD 1000 to 1900 are published in the National Geophysical Data Center (ftp://ftp.ngdc.noaa.gov/STP/SOLAR_DATA/AURORA).
- The index of north-south sunspot arrea assymetry *A* between AD 1821 and 1994, which is published in the Pulkovo observatory archive

The main methods , which are used there are the T-R periodogramm analysis for detecting of cycles in the time series (see Paper I ) as well as a multiple correlation-regressional analysis.

*3. The results and analysis*

*3.1 The temperature "residuals", middle latitude aurora and the north-south sunspots area assymetry*

As a next step in our study (see also the previous Paper I and II) it should be to estimate how is the possible contribution of the solar "erruptive" component in the studied temperature series . For the better discovering of the effect the correlation between the middle latitude aurora (MLA) annual numbers ($Aur\_N$) and the temperature "residual" series ($\Delta_2\Theta$) from the "sunspot activity –temperature" regressional model (4) in Paper I for the Northern hemisphere will be studied. As it has been mentioned in Paper II the using of MLA data is limited between AD 1700 and 1900 by two circumstances, namely: 1. The very possible serious lack of data before AD 1700; 2. The basic catalogue don't contain data for the 20$^{th}$ century.

It should to note that as a solar erruptive activity proxy the MLA annual number $Aur\_N$ has a disadvantage – it is very crude quantitative indicator of the total energy of these processes and consequently , of the penetrating in the Earth atmosphere high energetic solar corpuscular fluxes too. However there is no better proxy for these before the era of their instrumental observations.

The coincidience between the middle latitude auroral activity maximums and the subcenturial local temperature minimums as well as the opposite events too is shown on fig.1.

The all epochs of positive changes (warming) by mean duration of ~25-30 years are well corresponded to MLA fadding tendencies. In contrary, the cooling tendencies are dominating during the periods of the auroral activity increasing. As a result there is a well visible quasi- subcenturial (50-70 yr) "cooling-warming" cycle in approximately antiphase to the corresponding "auroral" cycle . Thus by fig.1 the conclusions for the reversed relationship between the high energetic solar corpuscular radiation and the Northern hemisphere temperatures is confirmed and visualized.

The coefficient of linear correlation between the both smoothed series is $r = -0.43$. It is about 7.5 times larger as its error and correspond to very high statistical significance (the "zero hypothesiss" probability there is $P < 10^{-6}$). The relationship is slightly better if a logarithmic type fitting for the relationship is used, i.e. $\Delta_2\Theta = a*ln(AurN) + b$. The correlation coefficient in this case is $-0.45$.

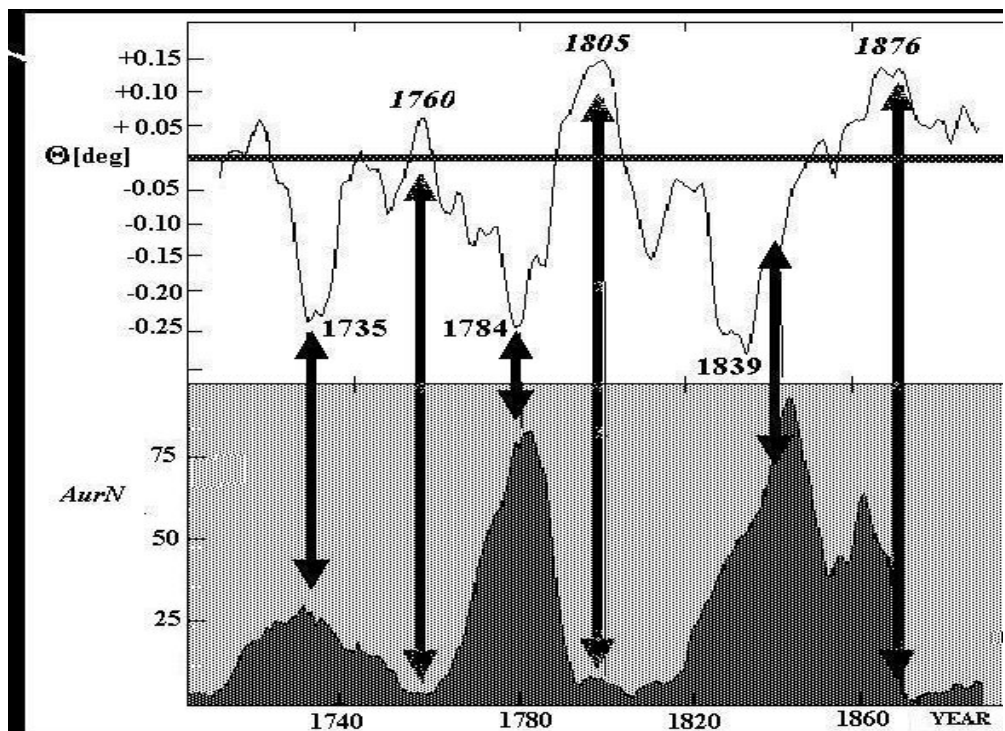

*Fig.1. Up: The the smoothing 11 year residual variations "$\Delta_2\Theta$" of Northern hemisphere temperature data (Moberg et.al.,2005) after the removing of the regressional model "Group sunspot numbers – Northern hemisphere temperature "(Paper I).The last one corrspond to the zero-level ((the dark horizontal line) ; Below : The smoothing 11 year annual MLA numbers.*

Because that the logarithmic relationship is less sensitive as the linear one it follow that the climate effect of the solar corpuscular events is significant if the activity of last ones is at enough high level. We have already note that the MLA annual number $Aur\_N$ is very rough proxy for this activity. It could not to estimate by $Au\_N$ how powerful are the separate MLA events , their corresponding solar sources and the corpuscular fluxes. This is why the so founded coefficient of correlation could be much better as the obtained above if a better proxy for the erruptive activity is used. Consequently the value of 20-25% from the total variance of the residual series, caused by the erruptive solar factor should be taken as the possible lower limit of the same one. Unfortunately there is no better proxy for the solar erruptive activity with enough long series before AD 1900 as the $Aur\_N$.

On other hand this relatively weak value of $r$ indicate that there a very important factor (or factors), most probably, is no taken into account. Is the last one connected to the Sun or it is by terrestrial origin?

Since the middle of 20[th] century there are a number of studies where the spatial distribution of the active regions on the solar disk as an important component of the

"Sun-climate" relationship is considerated. The most useful proxy for such aims is the index

$$A = \frac{S_N - S_S}{S_N + S_S}$$

(1)

where $S_N$ and $S_S$ are the total sunspot areas in the Northern and Southern hemisphere of the Sun respectively. As it has been pointed out by Loginov(1973) by geometric causes the Sun Northern hemisphere should essentially more geoeffective as the Southern one. As a "geoefficience" the ability of the solar erruptions to force over the Earth magnetosphere and atmosphere, causing geomagnetic storms, aurora, ionospheric disturbances and other geophysical events plus in the stratosphere and troposphere too is there in view off. Consequently, if there a climate forcing by solar erruptive events is assumed, the including of the assymetry index $A$ as an additional factor should be lead to much better model of the total solar effect over the "residual" series (the $\Delta_2\Theta$ -values), as if only the MLA annual number is taken into account. The new model should be of multi-factor type.

The best of sunspot arrea north-south assymetry data series for our aims is published in the Pulkovo Observatory Extended Data Archive. It contain the mean annual data of $A$-index since AD 1821 up to 1994. As in the case of the all other data the 11- year smoothed values are used there. The plot of the both smoothed series of the assimetry index $A$ and temperature residuals $\Delta_2\Theta$ (the dotted line) are shown on fig.2.

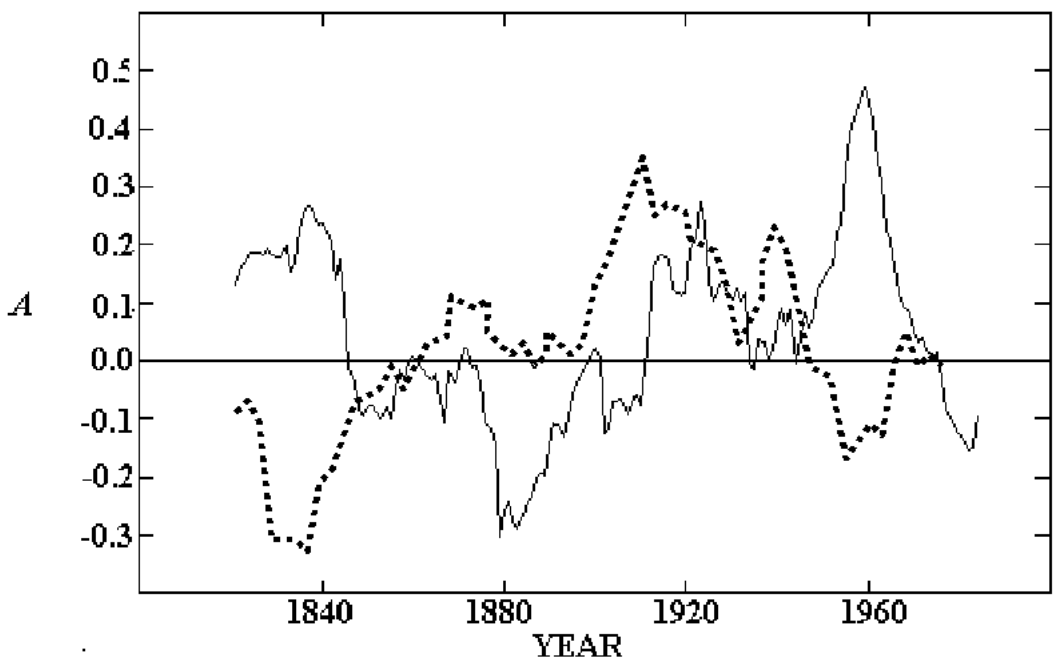

*Fig2. The north-south sunspot arrea assymetry index A during the period AD 1821-1994 (by the bold line) and the temperature "residual" $\Delta_2\Theta$ series in Celsius /Kelvin/ degrees (11-year smoothed values). The nummerical values on the Y-axis for the both series are identical.*

It is clear visible that the strongest negative values of $\Delta_2\Theta$ near to AD 1839-1840 and in the end of $1950^{th}$ are in very good coincidence with the local maximims of $A$, when the values of the last one are strong positive. The local warming maximum near to AD 1940 correspond to local minimum of $A$ too, but there the values of the last one are slightly > 0. There is only one period berween AD 1880-1910/1911 when the local cooling correpond to a weak local minimum of $A$. This period is interesting also by the last centurial solar minimum, which has been started at the end of $19^{th}$ century. But generally there is a well expressed anticorrelation between the north-south sunspot arrea assymetry index and the temperature changes in the Northern hemisphere of the Earth between the deepest phase of the solar Dalton minimum and AD 1980.

If the efficiency of the solar erruptive events over Earth climate depends by the north-south solar activity assymetry and the Northern hemisphere is essentially more geoeffective (Loginov, 1973), than the figures 1 and 2 are a good confirmation for the last one. The very deep local mimimum of of $\Delta_2\Theta$ near AD 1840 correspond well the both maximums of MLA and $A$ activity (fig,1 and 2). There are not catalogue data for the MLA in the middle of $1950^{th}$, but undoubtely the maximum of zurich cycle No 19 in AD 1957 correspond to a very high level of erruptive activiry. As it has been pointed out in Paper II the extrapolated maximums of MLA activity outside the end of Krivsky and Pejml catalogue data series should be near to AD 1910/1911 and 1975, while near to AD 1940 should be a minimum. This is well corresponding that the MLA activity during the cycles 18, 19 and 20 should be very high. By the combination with the strong maximum of $A$ (a very expressive domination of Sun Northern hemisphere activity) in $1950^{th}$ and remaining positive levels in $1960^{th}$ the temporal climate cooling between AD 1940 and 1975/76 could be satisfactorily explained.

This preliminary conclusion should be tested on the base of multi-factor correlation- regressional analysis. This part of the study has been provided on two stages.

The Northern hemisphere: AD 1821-1900

On the first stage the period before AD 1900 has been investigated. This separation is taken due to the fact that before this calendar year the MLA annual number $Aur\_N$ could be used as a proxy for the solar high energetic events. The smoothed 11-year data series of $A$, $Aur\_N$ and $Rh$ (the Group Sunspot Number) are used as possible factors for the changes of $\Delta_2\Theta$.

The first step there has been to determine the coefficients of linear correlation $r$ between the each pair of the parameters $\Delta_2\Theta$, $A$, $Aur\_N$ and $Rh$. They are correspondly: -0.407 for the pair $\Delta_2\Theta$ and $AurN$, -0.722 for $\Delta_2\Theta$ and $A$, and -0.560 for $\Delta_2\Theta$ and $Rh$.

The values of $r$ between the factors $Rh$ and $A$ should also to estimate. The higher by module values are an indicator that they are not enough independent each from other. It is very probable in many of such cases to exclude one of the both factors from the multiple model even if the coefficient of correlation between the factor and the predictant is high. The rule in this case is that in the model remain this factor which is better corellated with the predictant. The coefficients $r$ between the potencial factors are as followed:+0.530 for the pair $A$ and $Aur\_N$; + 0.621 for $Aur\_N$ and $Rh$, and +0.375 for $Rh$ and $A$.

The all obtained values of $r$ are statistically significant with probability >99.9%.

It need to point out the very high anticorrelation between the "residual" temperature data $\Delta_2\Theta$ and the sunspot arrea assymetry index $A$. On other hand there is a relative weak relationship between $Rh$ and $A$ and good correlation between $Rh$ and $\Delta_2\Theta$. The last one is better as the linear or logarythmic relationship between $Aur\_N$ and $\Delta_2\Theta$ ($r=$ -0.407). However the correlation between $Aur\_N$ and $Rh$ is higher ($r=+0.621$) as between $A$ and $Aur\_N$ ($r=+0.53$). These results shown that most probably in the multiple regression models the relationship between $\Delta_2\Theta$ and MLA could be totally captured and described by the terms describing *A* and *Rh* or their interraction .

A large number of one-, two- and three factor regressional models , including also different non-limear terms, has been obtained. The multiple coefficient of correlation *R* and the Snedekor-Fisher's *F*-test has been used for the selecting of the best of them. It has been found that the best of the all is:

$$\Delta_2\Theta = 0.07625 + 0.499*A - 1.224*A^2 - 1.943.10^{-8}*Rh^4 - 0.02334*A*Rh$$
$$R = 0.888; \quad F = 4.45 \tag{2}$$

The first important feature of this formula is the absence of any term, comtaining the annual number of middle latitude aurora , i.e *Aur_N*. The model (2) is a multiple function of two parameters - the group sunspot number *Rh* and the sunspot area assymetry index *A*. Obviously the influence of the solar high energetic corpuscules , for which the auroral activity index *Aur_N* as a proxy has used to this momennt , is better aproximated by the nonlinear terms of the types $Rh^4$ and $A*Rh$. Both coefficients of these terms are negative. It is provided by the $Rh^4$ term that there is a small climate cooling effect in the range of 0.2K , if the sunspot activity is very high (the smoothed 11-year *Rh* value is > 80- 100). The strong nonlinearity of this term expresses by our opinion the fact that during the investigated period the intensity of the most powerful erruptive events roughly correspond to the higher levels of sunspot activity.

The "interactial" two- factorial $A*Rh$ term is much more interesting. The sign of its climate effect depend by the sign of *A:* A negative value of the sunspot arrea assymetry correspond to a warming, while in the cases of higher sunspot activity in the Sun Northern hemisphere should be related to a climate cooling effect. If we use typical values for |*A*| =0.15 and *Rh*=50 the mean total amplitude effect over the $\Delta_2\Theta$ values is approximately 0.02334 x 0.15 x 50 x 2 = 0.35K. It expresses well the typical maximal deviations of the real smoothed 11-year Northern hemisphere and World Ocean temperatures to the corresponding "sunspot activity –temperature " models (Paper I). On this base it could argued that the "interactial" term gives in the most of cases the main yield for the $\Delta_2\Theta$ magnitude at least in the 19$^{th}$ century.

The last conclusion could be confirmed when an estimation of the both "pure" *A*-terms is made. The nonlinear $A^2$ -term gives always cooling effect, but it is significant only at high by module values of A (>0.2-0.3), while the linear term lead to cooling effect in range of 0.1K if *A*= - 0.2, or warming if *A*= +0.2. Thus the both "pure" *A*- terms are more important only during of the epochs of a very low sunspot activity when *Rh* tend to zero.

The *F*-parameter used there is defined as $F = S_t^2 / S_0^2$, where $S_t^2$ is the total variance (the factors variance $S_f^2$ +the residual variance $S_0^2$). Consequently for the model (2) $S_f^2 = 4.45 - 1 = 3.45$, i.e the both factors partition in the total variance is 3.45/4.45 =0.78 .Thus there is an evidence that 78% from the variations of $\Delta_2\Theta$ during the epoch AD 1821-1900 are by a certain solar origin and only 22%- by other not taken into account factors or ocasional data errors.

**The Northern hemisphere: AD 1913-1979**

As it has been already noted above, the catalog of Krivsky and Pejml ended at AD 1900 . On other hand it has been pointed out by our multiple regressional analysis that the index of assymetry and non-linear sunspot activity terms in the model are better proxy of the solar corpuscular events over the Northern hemisphere temperatures as the annual numbers of the middle latitude aurora. This is why for the multiple regressional analysis of the "residual" $\Delta_2\Theta$ temperature series during the 20$^{th}$ century only the *A* and *Rh* indicies as factors has been used. The aim is not only to estimate the yield of the solar corpuscular activity for the climate changes during the first 7-8 decades of the previous century, but also compare the potental evolution of the relationship with the same one during the 19$^{th}$ century.

It has been decide to choice as a start year AD 1913, because of the fact that by many authors has been started the new centurial solar cycle after the deepest phase of the Gleissberg-Gnevishev's centurial minimum (AD 1898-1923). This is the middle moment of the pair zurich cycles 14-15. On other hand near to this date is a breakpoint for the long time upward trend in the $\Delta_2\Theta$ data series, which has been started very soon after the end of the Dalton minima near to AD 1839/40.

The comparison of the pair coefficients of linear correlation $r$ show for three significant differences in relation to the period AD 1821-1900: 1. The coefficient of correlation for the pair $\Delta_2\Theta$ and $Rh$ is $r = -0.793$ vs $-0.560$ for the 19$^{th}$ century; 2. The corresponding coefficient is $r = -0.26$ between $A$ and $\Delta_2\Theta$ vs $-0.722$ (19$^{th}$ century), but it remain statistically significant at level >95%; 3. For the pair $A$ and $Rh$ the coefficient $r$ falls dramatically from $-0.375$ up to $-0.008$, i.e. unlike the 19$^{th}$ century a real relationship between the sunspots arrea north-south assymmetry and the sunspot activity is absent during the studied epoch. By our opinion the total independence of the both solar activity indicies each from other is the main cause for the changes of the other above mentioned relationships.

It has been found that the best from the all tested multiple regressional model is expressed by the formula:

$$\Delta_2\Theta = 0.715 - 0.965*A + 1.537*A^2 - 0.00852*Rh \qquad (3)$$
$$R = 0.861 ; F = 3.67$$

There is no "interactial" $A*Rh$ term in this formula, which is easy to explain- the smoothed assymetry index $A$ is positive in the almost all studied interval, except only one smoothed value near to AD 1933. It need to remember, that the yield of such term strongly depend no only by the sunspot actvity level, but also by in what Sun hemisphere this activity is predominated.

The factor variance in (3) is equal to $2.67/3.67 = 0.72$, i.e. about 72% from the total variance of $\Delta_2\Theta$.

The World Ocean "residual" series (AD 1856-1994)

The Pulkovo sunspot assymetry data series is ended at AD 1994. This is why the last AD 1995 has been excluded in the provided multiple regressional analysis for the World Ocean residual temperature data series. It need also to remember that the general statistical relationship "Group Sunspot Number – World Ocean temperature" is much closer as the corresponding one for the Northern hemisphere ($r = +0.877$, see formula (5) in Paper I). By the multiple regressional analysis procedure for the "residual" World Ocean temperature variations $\Delta_2\theta$ when $A$ and $Rh$ as factors (predictors) are used, it has been found that the best fitting is:

$$\Delta_2\theta = 4.535 - 0.462*A + 1.156*A^2 - 0.327*Rh + 0.08249*Rh^2 - 8.678 \cdot 10^{-5}Rh^3 + 3.24 \cdot 10^{-7}Rh^4$$
$$-0.0132*A*Rh$$
$$R = 0.814 ; F = 2.79 \qquad (4)$$

The model (4) is strong non-linear, but it contains the main features, which has been described above, i.e. general anticorrelation between the temperature "residuals" by one side and $A$, $Rh$ and $A*Rh$ by the other one. The factor variance $S_f^2$ is about 64% from the total variance $S_t^2$.

## 3.2. *The cycles in the sunspot arrea assymetry index A series (AD 1821- 1994)*

One of the steps of the present study is an analysis for an existence of cycles in the *A*- index data series. The main field of interests there is the possible existence of statistically significant cycles by subcenturial and nearcenturial duration and their comparison with the corresponding temperature "residual" data series spectra. On fig.3 the T-R correlogram (see Paper I) of the *A*-index series (AD 1821-1994) is shown. The time step $\Delta T$ is 0.5 years. The starting period is $T_0 = 2$ years, the upper limit is at $T=402$ years.

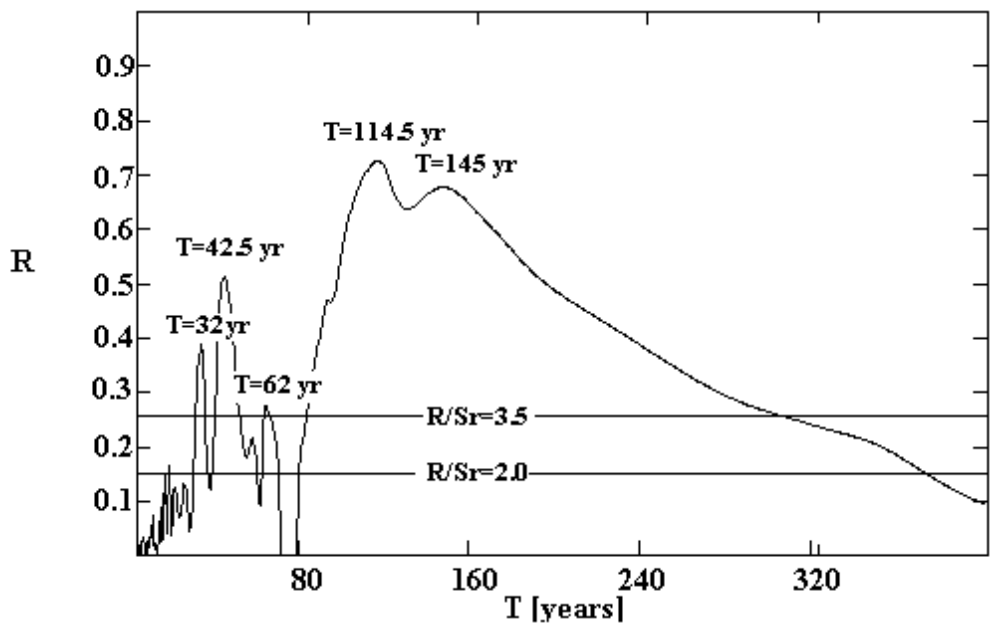

*Fig. 3 The T-R spectra of the north-south sunspot arrea assymetry index A (AD 1821-1994)(smoothed 11-year values)*

As it is shown there are two main cycles by durations of ~42.5 ( 4 Schwabe-Wolf's cycles) and 114.5 years. The second one is much more powerful. There is also an adjacent secondary peak at T=145 years, thus it could to assume that there is a doublette structure which mean duration is about 125-130 years.

However it should be taken this result with some reserve by the fact that the duration of the founded cycle is comparable with the length of the all data series (174 years), so by this one it could be determined rather as a "trend – hypercycle" as just a "cycle". On other hand the very high correlation coefficient *R*=0.72 at T=114.5 years show that most probable the quasi-120 year cycle is an important feature of the north-south sunspot arrea assymetry dynamics.

As it has been already shown in Paper II a hypercycle by duration of approximately 117 years has been found for the long-lived solar filaments (Duchlev, 2001). Conseqently, this is a variation of the solar corona dynamics, most probably not only for the filaments, but for other events (like CMEs) too. May be the north-south assymetry of the sunspot activity centers lead to long term tendencies of north-south anisotropy in the corona and the coronal events. This could affect corresponding GCR-flux fluctuations in the Earth atmosphere and on this base – fluctuations in the "cosmogenic" isothopes production rates , including the $^{10}$Be too. It is interesting in this course to mark also that a well expressed quasi –130-140 year cycle is visible also in the radiocarbon tree rings data series (Dergachev , 1994). Obviously the same phenomena should be affect the reaching the planets high energetic proton and electron fluxes (the solar cosmic rays, SCR) and this also is taken effect over their atmospheres and climate. The presence of this powerful quasi-120 year oscilation in the *A*-index data series gives by our opinion a principle explanation why a cycle with the same duration exist both in $^{10}$Be and climate data series, including also in the dendrochronological data (Komitov et al.,2003).

In fig.3 a weak , but statistically significant oscilation at T=62 year is shown. But as it has been already noted (see Paper I) there are weak oscilations by similar durations in the sunspot data series too (Komitov and Kaftan, 2003). On other hand there is a strong quasi- 60 year climatic cycle, which extremums are in antiphases with the corresponding ones of MLA series (see above), as well as with these of $^{10}$Be (Paper II) . These facts are a strong indicator for the solar corpuscular origin of the climatic 60-year cycle. However by our opinion it seems almost unpossible that the weak quasi-60 year cycles in *A* could be a source of the so powerful oscilations in the aurora, $^{10}$Be or climate. Obviously a much stronger solar source of this phenomena should be exist.

*3.3 The Group sunspot numbers (Rh) and their subcenturial (quasu-60 yr ) oscilations (AD 1610-1979)*

The multiple regressional analysis , which has been described above is clearly point out that the MLA annual number *Aur_N* have not any specific role as a factor proxy for the $\Delta_2 \Theta$ during the 19$^{th}$ century and ther participation as a factor is totally "captured" by *Rh* , *A* and the interactive term of the both last ones. On other hand a weak , but statistically significant 62 year cycle in the dynamics of the assymetry index *A* also exist . This is why the problem about the origin of the solar source of the subcenturial quasi 60 yr cycle has been rested open.

It has been very unexpetable when by using of the T-R periodogram procedure a strong quasi 67 year cycle for the epoch AD 1821-1900 has been detected in the smoothed 11 year *Rh* series (fig.4). As it is shown the correspondig corelllation coefficient *R* value is > 0.6 . In the T-R correlogram for the epoch AD 1913-1979 a totally dominant 59 yr cycle (*R* > 0.9; fig.5) is shown! No other traces of cycles in the subcenturial or quasicenturial range are visible in these both spectra. It is necessary to the studied time intervals are comparable by their length with the so detected subcenturial oscilations. By this one it is more correctly to determine the last ones as "trend – hypercycles", because they are occur only once in the both studied epochs.

These results are very intriguing, because usually for the overall sunspot activity where as a proxy the International Wolf's number is used, the cycles by longer duration (78, 88-90 , ~100 years) are discussed (Gleissberg, 1944; Vitinski et al., 1976 , Bonev, 1997). On other hand there have brief comments for a 65 yr cycle in the Schove's series (Schove, 1955) and weak variations in the subcenturial range (50-70 years) in the instrumental sunspot series *Rh* and *Ri* (Komitov and Kaftan,2003,2004) . This is why it has been decided to search how stable is this 60 yr cycle in the Group sunspot number series , as well as is there some evolution of the sunspot cycles in the centurial and subcenturial ranges during the period 1610-1979.

For this aim a two –dimensional T-R periodogram "moving window" procedure (MWTRPP, see Paper I) has been provided. The moving window lemght has been choiced to be 60 year and the parameters of the single T-R corellogram procedure are $T_0$= 2 years, time step $\Delta T$=0.5 years and $T_{max}$= 152 years correspondly. The evolution of the ratio *R/SR* (the ratio of *R* to its error) for the cycles in the range [$T_0, T_{max}$] is shown due to the map on fig.6.

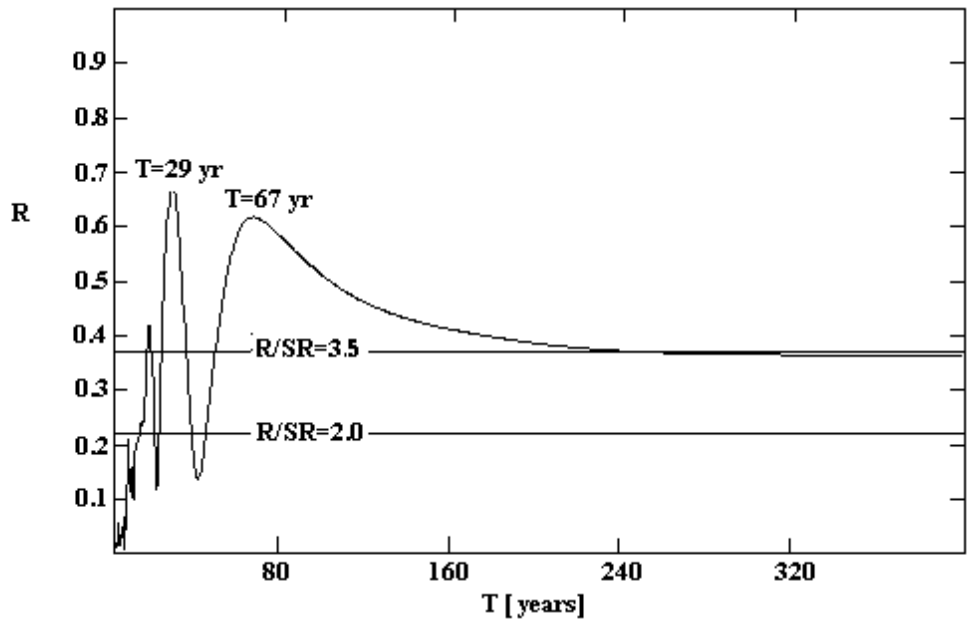

*Fig.4. The T-R spectra of the Group sunspot number (11-yr smoothing annual values; AD 1820-1900). Two powerful cycles by duration of 29 and 67 years are shown.*

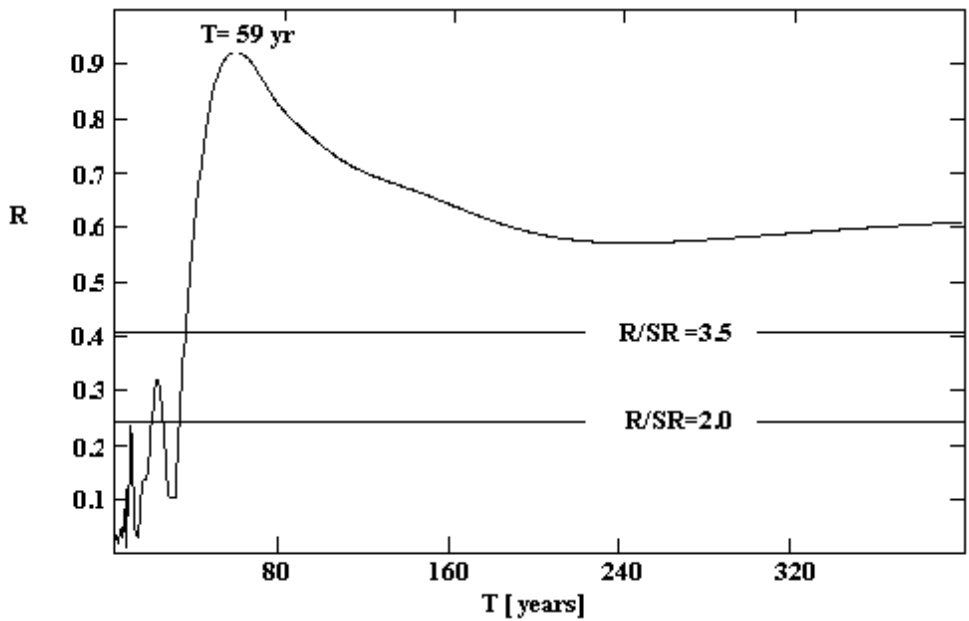

*Fig.5. The T-R spectra of the Group sunspot number (11-yr smoothing annual values; AD 1913-1979).*

The most interesting feature on fig.6 is related to the generally high level of presence of the subcenturial 50-70 yr oscilations in the *Rh* series before AD 1850. In the second half of 19$^{th}$ and the beginning of 20$^{th}$ centuries this type of cyclity has been sharply falling and this seems be much better expressed near to the zurich sunspot cycle No 13, i.e. before the Gleissberg-Gnevishev's solar centurial minimum. It is also clear visible a tendency for restoring of the quasi subcenturial cyclity after AD 1910. However as it is also shown there is slightly earlier also a tendency for longer quasi centurial 120 year trend -hypercycle . The last one is seems caused by the centurial minimum (AD 1898-1923). It is absent in the most recent "moving window" spectra after AD 1925-

1930 when the data from this minimum are not already included , i.e the last right columns of the map. A good confirmation for this is the strong peak at T= 59 years in fig.5 . In contrary the T-R spectra on fig.4 is related to an transition period from epoch with good expressed 60 yr cycle before AD 1850 and such one when  this cycle is totally absent.

Thus it could be say that the 50-70 year osilations are much more typical for the Group sunspot number data series during the last 400 years as every of  the non-stable quasi-centurial ones. It is even valid for the Maunder minimum epoch too, where the traces of subcenturial oscilations are weak , but even so , more visible as in the second half of 19$^{th}$ century.  It should also conclude that the abruption of  the 60 year cycle between AD 1850 and 1910 is the main cause by which it is not so visible in the general *Rh* series. It is necessary note that the second half of the 19$^{th}$ century is also a period of strong decreasing of the MLA activity, negative values of the *A*-index  and a fast climate warming (fig.1 and 2).

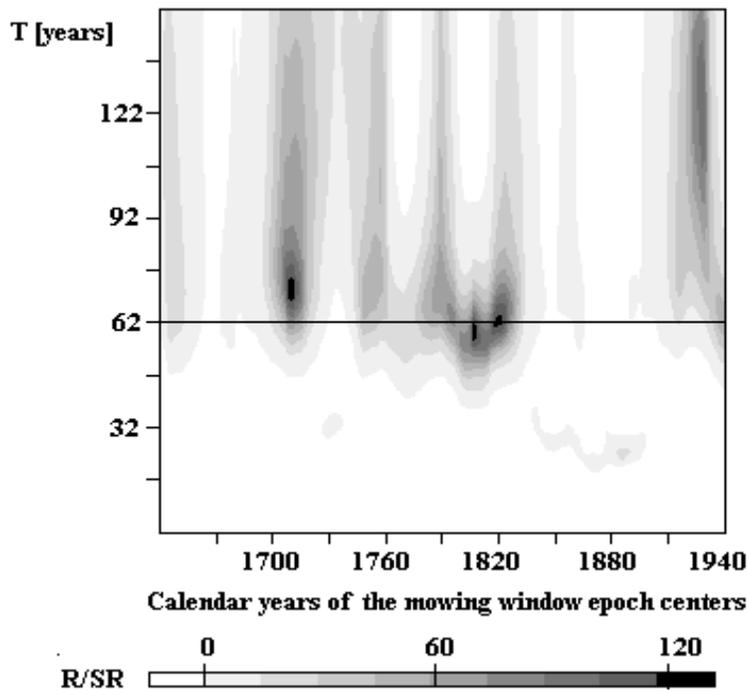

*Fig.6. The T-R spectra evolution of the 11 year smoothed GSN data in the range of periods T between 2 and 152 years. By the horisontal line of T=62 years the typical duration of the subcenturial cycle is signed. The most white arreas on the map to statistically non-significant values of  -0. 11 < R/SR < 0 are corresponded.*

There is no significant change of the results if the last 16 years up to AD 1995 of the Rh- series are included in the MWTRPP .

An comparison with  the Zurich series (the index *Ri*) could be very interesting, but it will be an object of a separate study.

*4. Discussion*

A few important conclusions should be derived by the presented in these three papers results and their analysis. There are also a number of questions , which remain still open or are new ones.

Most of all,  the general influence of the Sun over climate is much more rich and complicate as it is presented on the base only of  the TSI variations or even if the additional solar depended mechanism of galactic cosmic rays modulation over the aerosols and clouds production is taken into account. A third and very important Sun-climate forcing channel is related to the powerful erruptive events, which could generate high energetic protons (E ≥ 100 MeV). They are able to penetrate very deep in the Earth stratosphere and troposphere and even in the cases when E> 1 GeV to reach also the surface. A solar particles with such energies are labeled very often as solar cosmic rays (SCR).

The solar origin of MLA events is out of doubt. As it has been shown in Paper II for these phenomena during $18^{th}$ –$19^{th}$ century a powerful ~60 yr cycle is typical . The same one is valid for the Northern hemisphere temperature residuals and the "Greenland" $^{10}$Be series too. The analysis in Paper II as well as in this paper III are shown that these oscilation in the last two series are not by terrestrial, but rather by solar origin too. Most probably the $^{10}$Be and $\Delta_2\Theta$ 60 yr cycles are connected to the same solar phenomena as MLA. And there is a question – where on the Sun these phenomena are occurred? Are they coronal events like the coronal mass ejections (CME), or other components of the solar erruptive activity are also taken a significant participation in this channel of the solar forcing over the climate?

It need to say that there are not enough clear evidences on this stage about the dominant role of the coronal events (and especially of CMEs) for the generation of quasi- subcenturial climate as well as for the overall dynamics of the temperature "residual " series $\Delta_2\Theta$ and $\Delta_2\theta$ at all. It is rather visible by the results of the multiple regressional analysis both of the Northern hemisphere ($\Delta_2\Theta$) and the "oceanic" $\Delta_2\theta$ residual temperature series that the negative values (cooling) correspond in generally to an increasing of the overall sunspot activity index $Rh$ . It is an indicator that the cooling effect is related to the increasing effect of the erruptive events , which are close connected to the sunspot groups. This concern strongly especially the Northern hemisphere residuals $\Delta_2\Theta$ (formulas (2) and (3)) , while for the "oceanic" residual series the relationship with $Rh$ it is much more complicate (4).

The high amplitude and statistical significance of the quasi 60 year oscilations in the GSN data series is an additional evidence that both the climate and $^{10}$Be cycles with the same duration should be related to the overall erruptive activity. It also indirectly shown that the $Rh$ index is a very good proxy of the solar corpuscular activity. Unfortunately there are no any updates of the GSN data series after AD 1995. It could say on the base of the results in &3.3 that only by the relative short period at the end of $19^{th}$ century disturbs for the much better expression of the quasi-60 yr cycle in the GSN data series during the last 300 years after the Maunder minimum.

By our opinion the origin of the quasi –60 year cycle in the MLA events is now also clear. It is caused by the corresponding variations in the number of the active centers, which good proxy is the $Rh$ index. No any additional specific sources of solar activity for explanation of the 60 yr cycle of the auroral activity are strongly needed. So it is clear that our preliminary hypothesis for such sources (see Paper I and II) is not without fall necessary.

The obtained importance of 60 yr cycle in GSN data series lead on the top the question , what sunspot index is better – the international Wolf's number $Ri$ or $Rh$? The problem for a comparison of $Ri$ and $Rh$ as a proxies of the sunspot activity in different aspects of the solar-climatic relationships will be an object of our future paper. By the way there it will be only note that by opinion of many authors the $Rh$- index relative to the "classic" $Ri$ is much better proxy for the aims of solar-terrestrial physiscs at all , because it described much better the solar erruptive activity.

May be the most interesting result from this analysis is the obtained strong relationship between the Northern hemisphere temperature residuals and north –south assymetry index $A$ of the sunspot arreas. The last one is generally in not very strong relationship with the sunspot activity index $Rh$. By other words the $A$ – index is a second and relatively independent factor , which play very important role both for the $\Delta_2\Theta$ and $\Delta_2\theta$ dynamics. The linear correlation coefficent $r$ is equal to - 0.336 for the all period since AD 1821 (the beginning of the Pulkovo archive data) to 1979. It point out for a generall statistically significant reversed relationship. The coefficient $r$ is very high by module ($r = -0.72$) during the 19$^{th}$ century. For the recent part of $\Delta_2\Theta$ data series AD 1913-1979 the relationship has been sharp fadded ($r = -0.26$), but remain statistically significant over 95%.

The "interactive" terms ($Rh*A$) in formulas 2 and 4 shown that for the climatic effects over climate is very important no only the total level of the solar eruptive activity , but also where on the Sun the active regions are placed. The established fact that the relative increasing of the Sun Northern hemisphere activity lead to a climate cooling effects is in good agreement with the Loginov's suggestions about 35 years ago (Loginov, 1973), namely that the flare activity centers on the north of the Sun equator are more geoeffective as the southern ones.

As it is shown from the relationships (2-4) there are two "pure" north-south assymetry index terms of linear '$A$' and quadratic '$A^2$' types in the all three formulas. This is an indirect evidence that a climate forcing by solar activity processes, which are not very close connected to sunspot active centers should be also significant. Most probably these terms are connected directly to the coronal phenomena. It could be related no only to the CME events, but also to other coronal phenomena, including the more long time lived structures there and the large scale coronal structure too. It need to include in this course a possibility that the variations of the $A$-index could no only affect of the geoefficiency of the erruptive coronal processes like CMEs. They could also generate long time anisotropy variations of the interplanetary matter density and by this one lead to corrsponding time and space variations of the falling in the Earth atmosphere GCR –flux. As a result there should be effects over "cosmogenic" isothopes production rates, aerosols and clouds genetartion etc.

The index of the north south sunspots assymetry $A$ is an object of studying namely in the field of the solar –terrestrial and especially solar-climatic relationships (Georgieva, 2002). It is inetresting to point out for an interesting group of relationships between the north-south sunspot assymetry, the Earth diurnal rotation rate and the atmospheric circulation changes (Georgieva, 2002). The physical hypothesis for these influences is connected to the "solar wind –geomagnetic field – Earth dynamo" relationship. If this could be true and relationship between the sunspot flare proceses and sunspot assymetry from one side and the Earth tektonic activity from other should be exist. The possible relationship of the solar and Earth voulcanic activity is briefly discussed by the author in his recent overview (Komitov, 2008).

The multi-factor statistical models (2,3 and 4) pointed out that between 60 to 80% of the temeperature residuals $\Delta_2\Theta$ and $\Delta_2\theta$ are explained by the solar factors. The general anticorrelation both with the assymetry index $A$ and $Rh$, the very important participation of non-linear terms in these formulas as well as the "interactive" terms of type $A*Rh$ shown that these factors are predominantly connected to the solar corpuscular activity.

It need to remember that the general relationships "sunspot activity - temperature changes" ( formulas (4) and(5), Paper I ) express mainly the overall effect of the large time scale electromagnetic flux variations over the climate (the TSI changes). There could be also taken into account the possibility for a significant "hidden" participation in these models of the Forbush-effect and the GCR influence over the climate: The climate cooling effect of the GCR-flux increasing during the sunspot and TSI minimums should be made the overall cooling effect even more expressive. This is why the above mentioned models are explained very well the coincidience between the such significant phenomena like the solar supercenturial solar Maunder minimum (1640-1720) and the deepest phase of the last "Litle ice epoch", the next one ,also supercenturial Dalton minimum (1795-1835) and the temporal cooling during this time , as well as the Modern supercenturial solar minimum (1933-1996/2000) and the modern warm climate epoch.

However the climate effects of the solar erruptive activity are outside of these models. They are not taken into account in the most apropriated explanations for the climate changes in the modern epoch and especially during the last 35-40 years since the middle of 1970$^{st}$. The results and the analysis in our study are clearly pointed out that it is a very serious gap in the present domimamt climate changes theories. Only if the solar erruptive activity in combination with its spacial distribution over the Sun surface (the *A*-index) is taken into account it could be explain successfully the climate dynamics during the relative short "mirror epochs" ,when the sunspot activity and the temperature changes are in anticorrelation (see fig.1 in Paper I). This concern also the last 30-35 years. It will be demonstrated below how the specific combination of the solar erruptive activity and north –south sunspot area assymetry is the most probable factor for the fast warming both in the second half of 19$^{th}$ and in the end of 20$^{th}$ centuries and no additional cause (human activity) is needed for explanation there.

The solar activity and the climate changes during the 19$^{th}$ century

Undoubtly the most important solar activity event at the beginning of 19$^{th}$ century is the supercenturial Dalton minimum (AD 1795-1830/35) . The essential climate cooling during the this time is related to both the corresponding TSI decreasing and the increasing of the penetrating in Earth atmosphere GCR-flux. In generally this picture of the solar-climatic relationships during the Dalton minimum is correct. However there are some important details.

It is clearly shown on fig,1 that there is an initial period at the beginning of the Dalton minimum , when an increasing of the temperature residuals $\Delta_2\Theta$ in the Northern hemisphere is observed. The critical moment is near to AD 1805, i.e. near to or slightly after the maximum of zurich sunspot cycle No 5. So there is an delaying of about 10 years after the beginning of the Dalton minimum , when a sharp decreasing in the "residual" temperature series has been started. Consequently it should say that there is an additional climate cooling effect over the long –term downward TSI tendency. The sunspot activity during this time is low, and it is reflected to the low level of MLA activity (Paper II). Even so in the course of our results and analysis it should be assume that this cooling is caused by the solar erruptive factor and its increasing geoefficiency during this time. A very probable situation could be – relative rare, but strong erruptions , originated predominantly on the Northern hemisphere of the Sun. As it has been shown (fig.2) the Pulkovo archive data series is starting at AD 1821 with a positive values of the *A*-index . The tendency of *A* is positive during the next two decades and this correspond to even more deep cooling up to AD 1840, when the Dalton minimum has been already ended. However the *Rh* increasing after AD 1830 is predominantly in the Northern hemisphere and this has been supported the cooling tendency else certain time. A qualitative extrapolation of the smoothed *A*-index data in the past before AD 1820 shown that most probably it has been positive since 1805-1810, i.e well correspond to the observed decreasing of the temperature residuals since AD 1805.

Near to the maximum of zurich cycle No 9 a serious change in the long term solar activity tendencies has been occurred: 1. The "smoothed" assymetry *A*-index sign has been changed for long time (up to 1910-1912) from positive to negative. 2. There is a clear visible long term decreasing of the 11- year smoothed Group sunspot number data from zurich cycles No 9 to 14. The local peak near to AD 1870 is not affected seriously for this long time tendency (Paper I, fig.1) . Simultaneosly with these two events a long time $\Delta_2\Theta$ upward tendency from AD 1840 to 1910 has been occurred. There is only a short temporary stopping between AD 1870-1880. This dynamic of the climate changes is in very good agreement to the presented in this study results, their analysis and the following from them conclusions.

**The climate changes during the 20th century and the modern "global warming"**

After the centurial Gleissberg- Gnevishev's minimum (AD 1898-1923) the solar activity hast been very fast increased. After AD 1934 when sunspot cycle No 17 has been started and especially after 1940 the solar activity has been remained for a few decades on extremally high levels up to the end of sunspot zurich cycle No 22 in AD 1995/96 This epoch (the Modern supercenturial solar maximum ) is characterized no only with the higher for the last 1000 years Sun's luuminocity , but also with an extremally high erruptive activity which centers has been located predominantly in the Northern hemisphere (fig2.). As it has been shown there, the "smoothed" positive sign of $A$ has been remained up to the middle of $1970^{st}$. During the last 20 years after AD 1975 of the Pulkovo archive data series the sunspot assymetry index is predominatly negative. However the sunspot activity and the TSI index has been remained at high long-term level during this time.

As a result of all these circumstances , during the middle and the end of $20^{th}$ century is a supercenturial climate warming tendency maximum. It is connected to the supercenturial maximum of TSI on first place and the supercenturial GCR- flux minimum (less GCR –flux, less aerosols and clouds production ). However, on other hand the very high erruptive activity levels by itteraction with the positive $A$ –index lead to a secondary cooling effect. It is much better expressed between AD 1940 and 1975 when the flare activuty has been very fast increased. Since the $1970^{st}$ in coincidence with the transition of $A$-index from positive to negative the secondary cooling has been stopped and the general climate tendency has been changed to warming. The high levels of TSI during this time as well as the downward tendencies in the erruptive activity after the maximum of cycle No 21 (Komitov, 2008; Paper II) are additional factors to forcing of the warming during the last two decades of $20^{th}$ century.

**The zurich cycle No 24. What could be expected?**

There are many indications that with the end of solar cycle No 23 in 2008 a new supercenturial solar Dalton-type minimum is already started (Komitov and Bonev, 2001; Komitov and Kaftan 2003,2004; Shatten and Tobiska, 2003; Ogurtsov, 2005). This is why the next sunspot cycle No 24 should be essentialy weaker by magnitude as the previous few ones. The nearmaximal annual sunspot number $Ri$ in AD 2012 or 2013 is expected to be less than 100 , but there are also predictions for values near to or less than 50 (Cliverd et al., 2006). For an extremal low level of sunspot activity near to AD 2020 –2025 is poinetd out by Hathaway(2006) on the base of "Great Conveiyor Belt" model estimatons .

Consequently, a climatic significant decreasing of TSI during the cycle No 24 relative to No 23 as well as an increasing of GCR –flux should be expected as a general tendency. By this one and according the relationships between sunspot activity and climate a cooling effect in range of 0.2-0.3K if an decreasing of the 11 year smoothing $Rh$ values from 65-67 (at the beginning of $1990^{th}$ ) to 25 between 2010 and 2020 is assumed (see Paper I ,formula (4) and (5)). It need also to remember that there are not actual data for the Group sunspot number index $Rh$ after AD 1995 and any extrapolation should be made only on the base of some similarity with $Ri$.

However there should be added the effect of solar erruptions. Except a prediction for $Rh$ it need to have also a prediction for the index of sunspots area assymetry index $A$ too . If during the sunspot cycle No 24 the active centers are predominantly in the Northern hemisphere, which is by our opinion the most probable scenario (see also "Great Conveyor Belt" model results (Hathaway, 2006)), an additional cooling in order of 0.3 K for the Northern hemisphere over the above signed 0.25-0.3 K should be predicted. Thus a climate conditions could be returned back to almost the same ones as during the Dalton minimum.

There are historical evidences that the relationship between the auroral activity and the weather conditions in North Europa has been noticed still by the Vikings in the Middle Ages (Corbin, 2008). On other hand there are many studies during the 1950$^{th}$ - 1970$^{st}$ ,which are focused over the effects of the strong solar corpuscular erruptions for the climate. A good overviews of this studies is given by Rubashov (1963), Vitinskii et al.(1976) and Herman and Goldberg (1978). Our study shown that the understanding of this relationship is of most higher importance for the correct understanding of the climate changes in the modern epoch at all.